\documentstyle[emulateapj]{article}
\input epsf.tex
\newcommand{\be}{\begin{equation} }
\newcommand{\ee}{\end{equation}}
\begin{document}
\title{Halo White Dwarfs, Thick Disks and a Sanity Check}
\author{Brad M. S. Hansen }
\affil{ Hubble Fellow \\
Department of Astrophysical Sciences, Peyton Hall, Princeton University,
Princeton, NJ, 08544}
\authoremail{hansen@astro.princeton.edu}

\begin{abstract}
The discovery of a population of high proper motion white dwarfs by
Oppenheimer et al (2001) has caused a lot of speculation as to the
origin of these stars. I show that the age distribution of the white
dwarfs offers a kind of sanity check in these discussions. In particular,
this population appears to have a similar age distribution to those in
the standard, thin disk white dwarf population. This is not what is
expected for either the halo or thick disk, which are thought to
be old populations. It may indicate a different dynamical origin entirely,
or it may be just be the tail of a larger distribution, implying a 
significantly larger total density in even older white dwarfs.
\end{abstract}


\section{Introduction}

The recent announcement of the detection of a significant population of
old white dwarfs with high proper motions (Oppenheimer et al 2001; hereafter OHDHS) has 
sparked a lot of interest in the nature of this population. If this
population is indeed representative of the Galactic halo, as claimed by
OHDHS, then it could represent a significant contribution to the claimed
microlensing population (Alcock et al 2000). However, the assignment of these
white dwarfs to a particular dynamical family is still somewhat controversial.
In particular, Reid, Sahu \& Hawley (2001; hereafter RSH), by comparing
with a local sample of M~dwarfs, suggest that this
population is actually representative of the thick disk and consistent with
known Galactic populations. In either case, the OHDHS sample represents a
new and interesting probe of the old stellar populations in the solar neighbourhood.

Most of the recent controversy regarding the OHDHS claim centers on the dynamical
interpretation of their results; i.e. whether the detected white dwarfs represent
the true halo or the thick disk and whether the density required is consistent with
the expected stellar populations or requires an interpretation in terms of a
`dark matter' contribution. In this article I wish to address another aspect of this
population, namely the age distribution and what that can tell us about the origin
of these white dwarfs.
First of all, in \S\ref{ThickDisk} I will review the arguments about the various
dynamical populations suggested and the kinds of densities expected therefrom.
 Therafter, I will
 discuss in \S\ref{WDLF} the age distribution of the OHDHS sample and will
discuss the implications in
 \S\ref{Conc}.

\section{Thick Disks \& Halos}
\label{ThickDisk}

The OHDHS `halo' sample of white dwarfs is culled from an area covering
about 4000 square degrees towards the South Galactic Pole, or about $\sim 10\%$ of
the sky. The magnitude limit of the search is $R=19.8$, while the proper motion 
limit is $0.33''\rm .yr^{-1}$. They apply a further velocity cut, selecting only
those white dwarfs with velocity $V_{\perp}>94\, {\rm km.s^{-1}}$ (using photometric
distance estimates) in order to define their high-velocity `halo' population.
Using a V$_{\rm max}$ analysis they infer a mass density
 $\sim 1.3 \times 10^{-4} \rm M_{\odot}.pc^{-3}$ in this high velocity population.

This density is considerably larger than the estimate of $1.3 \times 10^{-5} \rm M_{\odot}.pc^{-3}$ 
expected from the standard spheroid (Gould, Flynn \& Bahcall 1998) and prompts interpretations
in support of halo white dwarf populations from microlensing (e.g. Alcock et al 2000).
However, alternative explanations have already been suggested. RSH have shown that
a similar velocity cut on their volume-complete sample of M~dwarfs from Reid, Hawley \& Gizis (1995) 
allows 20 out of 514 stars ($\sim 4\%$) into the `halo' sample thus defined. As such, they
interpret the OHDHS white dwarfs as simply a high-velocity tail of a disk population. This
is also supported by the fact that the OHDHS `halo' sample does show a distinctly
asymmetric velocity distribution, suggesting some fraction belongs to a rotating
component. 

To understand the various positions on this issue, we need to understand what is
expected from each population. The thick disk is generally defined 
 as a co-rotating (with the sun) population with vertical scale height
$\sim \rm 1 kpc$, vertical velocity dispersion $\sim 40-45\, {\rm km.s^{-1}}$ and a number density
several percent of the local disk value (Sandage \& Fouts 1987; Reid \& Majewski 1994; Robin 1994).
We should note, at this point, that the M-dwarf sample of Reid, Hawley \& Gizis (1995) seems
to contain a significantly larger fraction in the high velocity population ($\sim 15-20\%$).

To infer absolute numbers, we review the range of 
Galactic mass models from Dehnen \& Binney (1998). They assume a 5\% fraction of the total disk mass
in the thick component and find a range of {\em total} thick disk mass
 ranging from
 $8.3 \times 10^{-4} {\rm  M_{\odot}.pc^{-3}}$
to $ 1.11 \times 10^{-3} {\rm M_{\odot}.pc^{-3}}$.
 However, applying the velocity cut at $94 \, {\rm km.s^{-1}}$ removes all but the highest velocity objects
from the sample. OHDHS claim this as a $2 \sigma$ cut. In reality, applying this cut to the proper motions
drawn from a 3-D maxwellian only removes $\sim 87\%$ of the thick disk stars. Furthermore, allowing for a
$\sim 35\, \rm km.s^{-1}$ lag in the thick disk rotation velocity (Chiba \& Beers 2000), suggests as much
as $20 \%$ of the stars make it into the sample. Thus, we may expect a range
$\sim 9.1 \times 10^{-5} {\rm M_{\odot}.pc^{-3}}$--$2.2 \times 10^{-4} {\rm M_{\odot}.pc^{-3}}$. Hence the
white dwarf density inferred by OHDHS appears to be consistent with these numbers. However, one
potentially disturbing feature of this agreement is that we have yet to account for the
mass in stars below the turnoff i.e. those which have yet to form white dwarfs.

To determine whether this is indeed disturbing, we have to infer the fraction of the total population
mass that we expect to reside in white dwarfs. Assuming a Salpeter function from 0.1~$M_{\odot}$ to
$8 M_{\odot}$ and assuming all stars with $M>M_{TO} = 0.82 M_{\odot}$ form $0.6 M_{\odot}$ white
dwarfs, the ratio of mass in white dwarfs to mass in stars with $M<0.82 M_{\odot}$ is
\begin{eqnarray}
\frac{M_{wd}}{M_{sub-TO}} & = & \frac{ 0.6 M_{\odot} \int_{M_{TO}}^{8 M_{\odot}} M^{-(1+x)} dM}
{ \int_{0.1 M_{\odot}}^{M_{TO}} M^{-x} dM} \\
 & = & \frac{x-1}{x}0.6 M_{\odot} \frac{M_{TO}^{-x}-\left( 8 M_{\odot} \right)^{-x}}
{ \left( 0.1 M_{\odot} \right)^{1-x} - M_{TO}^{1-x}} \\
 & = & 0.14 \, \left( {\rm M_{TO}=0.82 M_{\odot};\, x=1.35} \right)
\end{eqnarray}
so that $\sim 13 \%$ of the total stellar mass is contained in white dwarfs. Thus, the expected
total mass density in local, thick disk white dwarfs above the velocity cut is  $< 2.8 \times 10^{-5} {\rm M_{\odot}.pc^{-3}}$,
considerably less than the OHDHS determination (and similar to the Gould et al number for the spheroid).
Of course, the Salpeter mass function diverges at the low mass end, so this number could
be sensitive to the low mass cutoff. Thus, let us repeat this calculation 
using the empirical disk mass function of Gould, Bahcall \& Flynn (1996) (and which doesn't
diverge at the low mass end).
Instead of a Salpeter slope x=1.35, this mass function has x=1.21 (note the different sign
convention from Gould et al) between $0.6-0.73 M_{\odot}$
and $x=-0.44$ for $M<0.6 M_{\odot}$. Gould et al argue that a correction for unresolved
binaries will increase $x$ to $x \sim 0$, so we adopt $x=0$ at the low end.
  Using this mass function (extrapolated
through the white dwarf region), the ratio of
mass in white dwarfs to sub-turn-off stars is 0.41, i.e. a 29\% share of the total mass budget. Thus,
a more conservative estimate is $6.8 \times 10^{-5} {\rm M_{\odot}.pc^{-3}}$  thick disk white dwarfs in the
OHDHS sample, i.e. only a factor of two smaller than the OHDHS number.

Of course, there are several assumptions in such a naive model (mass function extrapolations, simple
maxwellian velocity distributions) which can be changed to provide better agreement. We will consider
these issues in \S\ref{Conc}, but first we consider the other curious feature of the OHDHS sample,
namely the age distribution.

\section{The Thick Disk White Dwarf Luminosity Function}
\label{WDLF}

Again, it is useful to understand what we expect from a simple model. The thick
disk is kinematically and chemically distinct (e.g. Freeman 1993; Majewski 1993) from the thin disk and is thought
to be a population that formed primarily in a burst $\sim$12~Gyr ago (e.g. Gilmore, Wyse \& Jones 1995).
Thus, we model this as a single burst of star formation 12~Gyr old, whose white dwarf
ages vary depending on the mass and consequent main sequence lifetime of the progenitor.
To make a conservative model, we
 want to maximise the fraction
of the white dwarf population that would be detectable by OHDHS.
We assume a standard Salpeter IMF, a main sequence lifetime
based on the models of Hurley, Pols \& Tout (2000) 
 and we shall assume that all stars above the turnoff mass 
(0.82~$M_{\odot}$ in this model) make $0.5 M_{\odot}$ white dwarfs (more massive
white dwarfs cool more rapidly at late times because of earlier core crystallisation), with hydrogen
atmospheres (because these are the slowest cooling) and pure Carbon cores (because these have
the largest plausible heat capacity and thus cool most slowly).
 Thus, we are skewing the white dwarf luminosity function to
the bright end.

\begin{figure}
\plotone{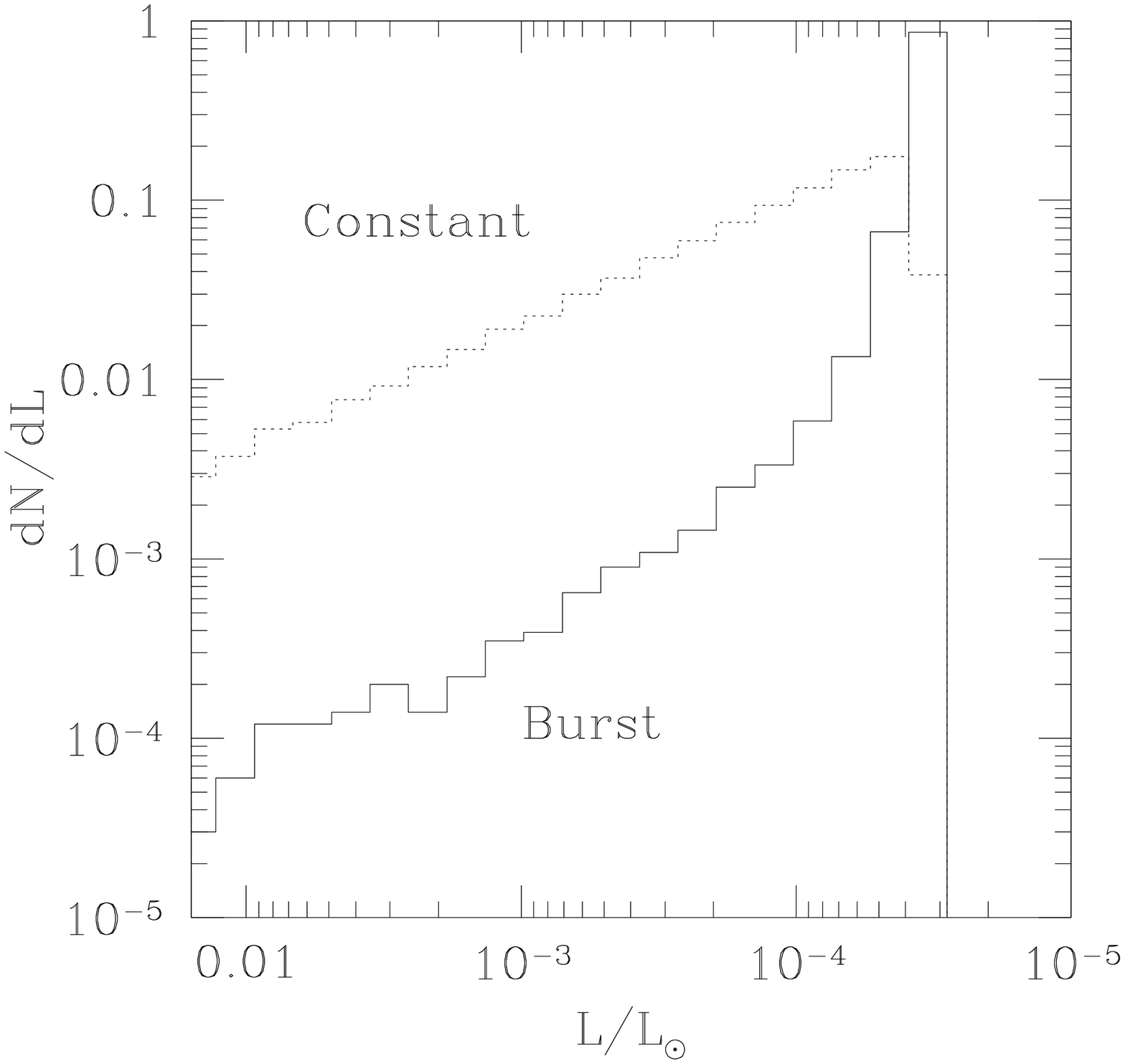}
\caption{ The lower solid curve is the distribution of white dwarf luminosities, assuming
the model described in the text, for a single burst of star formation 12~Gyr ago. The
upper dotted line is the distribution using the same model but assuming a constant star
formation rate. The two distributions are normalised to have the same total number of stars
formed.\label{ill}}
\end{figure}

Figure~\ref{ill} shows the resulting white dwarf luminosity function, using the models of
Hansen (1999). For comparison we 
include a luminosity function with all the same input parameters except that we assume
a constant star formation rate over the last 12~Gyr. The most striking difference is
that the burst population has a much sharper rise at the faint end. Thus, the vast
majority ($\sim 90 \%$) of the white dwarfs lie within 0.5 magnitude of the faintest white
dwarfs in the burst case. This difference in scalings also demonstrates that one cannot
 simply try to scale the ratio of white dwarfs to M-dwarfs from the thin disk
(a $\sim$ constant star formation rate population) to the thick disk (if it is a burst population)
as done in RSH. 

The next question to ask is whether the OHDHS white dwarfs are indeed old enough to
account for the thick disk population. Again, we use our most conservative, 0.5~$M_{\odot}$,
Carbon core, Hydrogen envelope models. Thus, we derive upper limits on the age. 
The absolute R-band magnitudes span the range $\rm M_R=13-16$.
Using our conservative
model, one finds an age of $\sim 6.5$~Gyr for $\rm M_R=15$ and $\sim 10.4$~Gyr for $\rm M_R=16$. An age of 12~Gyr corresponds
to $\rm M_R\sim 16.5$, i.e. the oldest white dwarfs in the sample are not older than $10$~Gyr and most are considerably
younger. Furthermore, recall that these are the most conservative models, i.e. the ages are probably younger
than this.
Obviously the expected age of 12~Gyr is the combination of both main sequence and white dwarf lifetimes,
but, as we have shown above, we expect the white dwarfs to pile up at the faint end in a population
resulting from a burst. Another way of putting this is to infer the main sequence mass of the progenitor
by subtracting the inferred white dwarf age from the presumed 12~Gyr age of the burst. The faintest white
dwarf ($\rm M_R \sim 15.7$) has a white dwarf age $<9.9$~Gyr (recall we are using the slowest cooling models)
and thus comes from a star less massive than $\rm 1.35 M_{\odot}$ if originating from a 12~Gyr burst. However,
a more typical representative of the coolest OHDHS white dwarfs has $\rm M_R \sim 15.1$, an age $\sim 6.9$~Gyr and
a progenitor mass $\sim \rm 0.96 M_{\odot}$. If we adopt the Gould et al mass slope for the burst we find that (extrapolating to progenitor
masses of 8$ \rm M_{\odot}$), we are missing the white dwarfs from between 50\% and 80\% (depending on the
completeness of the OHDHS sample) of the total population.
Since the observed densities are already slightly discrepant with our expectations, such
 enhancements of the density by factors of 2--5 would lead to a significant discrepancy.

How much of this discrepancy could be due to inaccuracies in the cooling models? Intercomparisons of
different theoretical groups show model ages which differ by $\sim$ 10\%, not enough to alleviate the
above comparisons (a good review of the current state of the field can be found in 
Fontaine, Brassard \& Bergeron (2001)).
In particular, Fontaine et al present a model (pure Carbon core, Hydrogen atmosphere) similar to that of
our above conservative model (although the mass is $\rm 0.6 M_{\odot}$). We can compare the luminosity of
such a 12~Gyr white dwarf from the two codes (and, of course, using a $\rm 0.6 M_{\odot}$ model from our
tables) and find $\rm \log L/L_{\odot} = -4.73$ (Fontaine) and $\rm \log L/L_{\odot} = -4.78$ (Hansen). 
The above white dwarfs are considerably brighter and younger.

\begin{figure}
\plotone{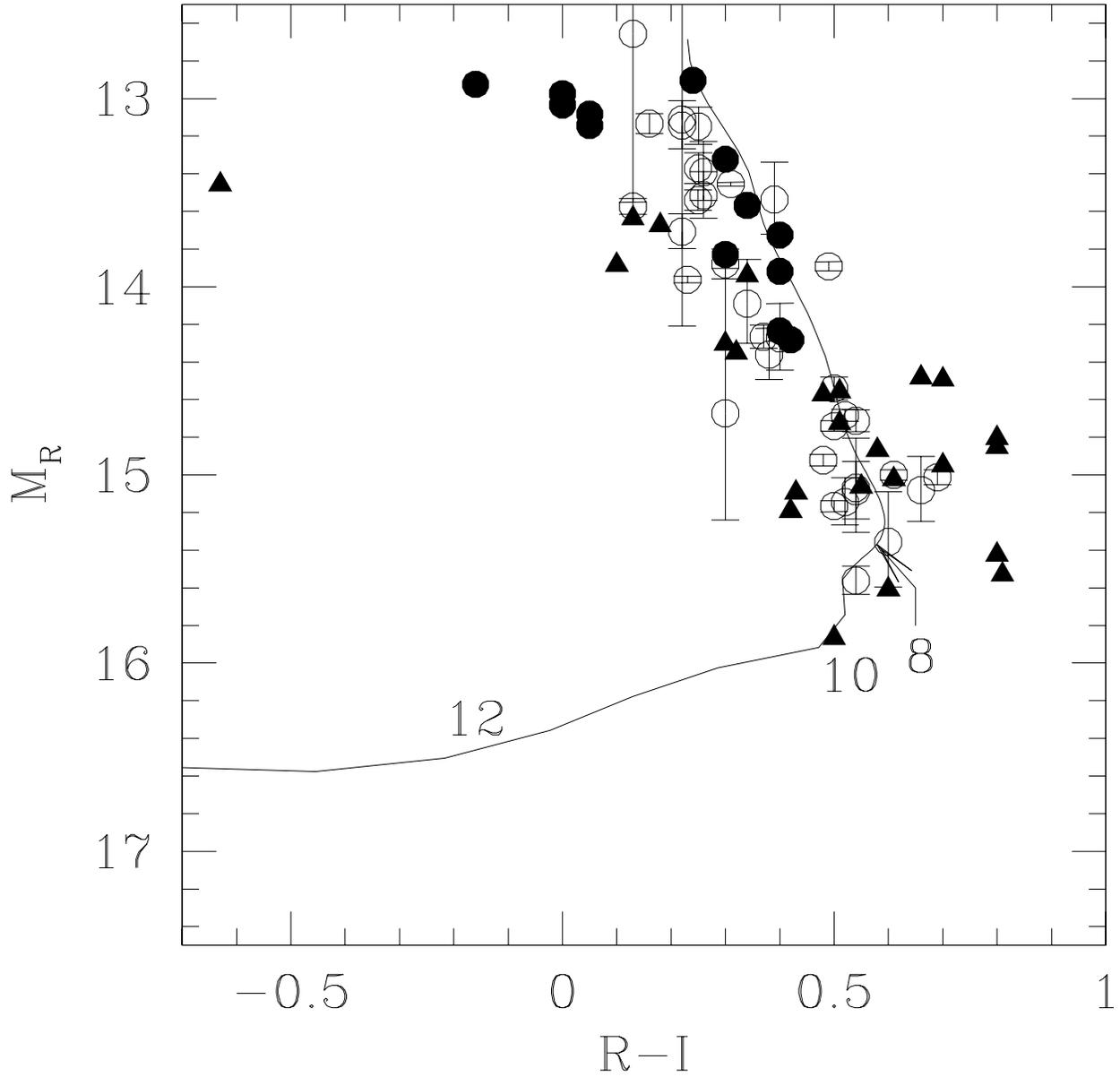}
\caption{ We show here the absolute magnitudes (the 20\% error bars on the photometric distances
are not shown).
 The solid points are the OHDHS dwarfs (circles show $\rm H\alpha$ and triangles do not)
 and the open points are the thin disk dwarfs from Bergeron et al. The
solid curve shows a 0.5$\rm M_{\odot}$ pure Carbon core model with a Hydrogen atmosphere.
Note that this is not intended as a fit to the observed sample but represents
the slowest cooling white dwarf plausible. The positions of white dwarfs of ages 8, 10 and 12~Gyr are
shown.\label{RI}}
\end{figure}

The most empirical measure of the ages of the OHDHS white dwarfs is
 to compare them directly 
to the thin disk sample of Liebert, Dahn \& Monet (1988) ( although we will use the
photometry of Bergeron, Ruiz \& Leggett (1998)). Figure~\ref{RI} shows that most of the
OHDHS white dwarfs have similar ages to the thin disk white dwarfs, and a similar
distribution. On the basis of this diagram alone, it would be difficult to distinguish this
new population from the standard thin disk white dwarf population.

Given that the OHDHS white dwarfs appear brighter than expected for their proposed parent
population, one might wonder if there is some gross error in the photometric distance indicator
they used. The claimed 20\% error is probably reasonable, given the small variation in radius
between white dwarfs of different mass and composition. One potential note of caution though,
is that mixed Hydrogen and Helium atmospheres can wreak havoc with the colours of cool white
dwarfs (e.g. the non-monotonic behaviour of the mixed model shown in OHDHS Figure 4). This
could contribute some additional systematic uncertainties. However, a wholesale overestimate
of the distances by this method is probably unlikely, given that many of the white dwarfs show
$\rm H\alpha$ emission and the edge of that subset lies at an inferred $\rm M_R \sim 14.5$ corresponding
to temperatures $\sim 5000$~K (where we indeed expect the $\rm H\alpha$ to become undetectable).


Finally, given these interesting results, one can ask what is required to see the kind of old white
dwarfs we expect. Searches in the R-band are optimal for the colours of old white dwarfs (Hansen 1998, 1999;
Saumon \& Jacobsen 1999), and even the slowest cooling models described above predict $M_R \sim 16.5$ for
a 12~Gyr white dwarf. To distinguish thick disk/halo white dwarfs we also require $\rm V_{\perp} > 100 km.s^{-1}$.
Thus, we may set a target reduced proper motion 
$\rm  H_R   =  23.1 + 5 \log \left( V_{\perp}/100 km.s^{-1} \right) $. The OHDHS sample is not complete
at this level and thus there may still be a lot of white dwarfs to be discovered.

\section{Conclusions}
\label{Conc}

Based on the considerations of the previous section, it is tempting to conclude that the white dwarfs
identified by OHDHS have a provenance similar to those in the thin disk. Certainly, on the basis of
the colour-magnitude distribution alone, it is difficult to distinguish the two populations. However,
this sample was drawn from a high velocity sample. While arguments rage about their membership
in the halo or the thick disk, kinematic membership in the usual thin disk population seems unlikely.
Of course, underlying this is the assumption that even old thin disk populations are described by
maxwellian velocity distributions with dispersions $\rm \sim 20 km.s^{-1}$. Perhaps this is simply
indicating that there are other dynamical processes at work in the disk which can pump some stars up to 
 velocities characteristic of the thick disk.

The thick disk is generally regarded as kinematically and also largely chemically distinct from
the old thin disk (Gilmore \& Reid 1983; Sandage \& Fouts 1987; Gilmore, Wyse \& Kuijken 1989; Freeman 1993, Reid \& Majewski 1993).
 If the OHDHS white dwarfs are truly part of the thick disk they rather belie
this distinction, suggesting that perhaps star formation in the thick disk is more complex than
a simple burst at early times.
One can make the OHDHS sample consistent with our preconceptions about the thick disk if one believes
that it represents only the bright tail of a larger distribution. However,
given the inferred densities as described in section~\ref{ThickDisk}, a further addition of significant
mass would again imply a thick disk white dwarf contribution considerably larger than previously thought.
As such, it would imply a top-heavy mass function i.e. one more weighted towards the production of white
dwarfs than a standard Salpeter mass function. Our naive estimates of \S\ref{WDLF}  
suggest as much as 80 \% of the white dwarfs could lie beyond the OHDHS limits. Note, however, that
the increased mass in white dwarfs would still not dominate the local disk mass, since the addition
would only be of order the currently accepted (total) thick disk mass i.e. still an order of magnitude less than
the thin disk contribution.

Finally, much of the controversy regarding the interpretation of the OHDHS observations is motivated
by whether this represents a detection of the population of stars responsible for the microlensing 
observed by Alcock et al (2000). For this purpose it is not sufficient to distinguish between true
halo and thick disk populations, as there are models (Gyuk \& Gates 1999) which attempt to explain
the microlensing results using rotating populations, reminiscent of a very thick disk. Certainly,
if one were to find significantly more white dwarfs (be they halo or thick disk) at even fainter 
magnitudes, the mass required
would begin to point towards such a population.

This work was supported by NASA through Hubble Fellowship grant HF-01120.01-99A from the Space
Telescope Science Institute, which is operated by the Association of Universities for Research
in Astronomy, Inc. under NASA contract NAS5-26555.

\clearpage


\end{document}